%
%
%
%
%
%
%
%
%
%
%
%
\hoffset=0.0in
\voffset=0.0in
\hsize=6.5in
\vsize=8.9in
\normalbaselineskip=12pt
\normalbaselines
\topskip=\baselineskip
\parindent=15pt
%
%
%

\let\dl=\delta

\let\la=\langle
\let\ra=\rangle
\let\pa=\partial
\let\lf=\left
\let\rt=\right
\let\dt=\cdot
\let\del=\nabla
\let\dg=\dagger

\let\q=\widehat

\let\h=\hbar

\let\rta=\rightarrow

\let\x=\times

\let\sy=\scriptstyle

\let\:=\>
\let\\=\cr
\let\emph=\e

\let\m=\hbox

\let\cl=\centerline

\def\e#1{{\it #1\/}}
\def\textbf#1{{\bf #1}}
\def\[{$$}
\def\]{\[}
\def\re#1#2{$$\matrix{#1\cr}\eqno{(#2)}$$}
\def\de#1{$$\matrix{#1\cr}$$}

\def\eqdf{\buildrel{\rm def}\over =}
\def\hf{{\sy {1 \over 2}}}
\def\qr{{\sy {1 \over 4}}}

\def\qH{\q{H}}

\def\mathrm#1{{\rm #1}}

\def\mathcal#1{{\cal #1}}

\def\mbf{\fam\bffam\tenbf}
\def\bv#1{{\mbf #1}}

\def\vr{\bv{r}}

\def\vp{\bv{p}}

\def\vA{\bv{A}}

\def\vj{\bv{j}}
\def\vk{\bv{k}}

\def\vPhi{\bv{\Phi}}
\def\vPi{\bv{\Pi}}
\def\vPsi{\bv{\Psi}}

\def\qvp{\q{\vp}}

\def\phs{\phi_\psi}
\def\pis{\pi_\psi}
\font\frtbf = cmbx12 scaled \magstep1
\font\twlbf = cmbx12
\font\ninbf = cmbx9
\font\svtrm = cmr17
\font\twlrm = cmr12
\font\ninrm = cmr9

\def\abstract#1{{\ninbf\cl{Abstract}}\medskip
\openup -0.1\baselineskip
{\ninrm\leftskip=2pc\rightskip=2pc #1\par}
\normalbaselines}
\def\sct#1{\vskip 1.33\baselineskip\noindent{\twlbf #1}\medskip}

\def\so{\raise 0.65ex \m{\sevenrm 1}}
\def\sk{\par\vskip 0.66\baselineskip}
{\svtrm
\cl{Equivalence of classical Klein-Gordon field theory}
\medskip
\cl{to correspondence-principle first quantization of}
\medskip
\cl{the spinless relativistic free particle}
}
\bigskip
{\twlrm
\cl{Steven Kenneth Kauffmann}
}
\cl{American Physical Society Senior Life Member}
\medskip
\cl{43 Bedok Road}
\cl{\#01-11}
\cl{Country Park Condominium}
\cl{Singapore 469564}
\cl{Tel \&\ FAX: +65 6243 6334}
\cl{Handphone: +65 9370 6583}
\smallskip
\cl{and}
\smallskip
\cl{Unit 802, Reflection on the Sea}
\cl{120 Marine Parade}
\cl{Coolangatta QLD 4225}
\cl{Australia}
\cl{Tel/FAX: +61 7 5536 7235}
\cl{Mobile:  +61 4 0567 9058}
\smallskip
\cl{Email: SKKauffmann@gmail.com}
\bigskip\smallskip
\abstract{
It has recently been shown that the classical electric and magnetic fields
which satisfy the source-free Maxwell equations can be linearly mapped into the
real and imaginary parts of a transverse-vector wave function which in
consequence satisfies the time-dependent Schr\"{o}dinger equation whose
Hamiltonian operator is physically appropriate to the free photon.  The
free-particle Klein-Gordon equation for scalar fields modestly extends the
classical wave equation via a mass term. It is physically untenable for
complex-valued wave functions, but has a sound nonnegative conserved-energy
functional when it is restricted to real-valued classical fields.  Canonical
Hamiltonization and a further canonical transformation maps the real-valued
classical Klein-Gordon field and its canonical conjugate into the real and
imaginary parts of a scalar wave function (within a constant factor) which in
consequence satisfies the time-dependent Schr\"{o}dinger equation whose
Hamiltonian operator has the natural correspondence-principle relativistic
square-root form for a free particle, with a mass that matches the Klein-Gordon
field theory's mass term.  Quantization of the real-valued classical
Klein-Gordon field is thus second quantization of this natural
correspondence-principle first-quantized relativistic Schr\"{o}dinger equation.
Source-free electromagnetism is treated in a parallel manner, but with the
classical scalar Klein-Gordon field replaced by a transverse vector potential
that satisfies the classical wave equation.  This reproduces the previous
first-quantized results that were based on Maxwell's source-free electric and
magnetic field equations.
}

\sct{Introduction}
\noindent
The classical Hamiltonian for the spinless relativistic free particle
is $(|c\vp|^2 + m^2c^4)^\hf$, which, from the correspondence princi%
ple, unequivocally implies that its first-quantized description is
via state vectors $|\psi(t)\ra$ that satisfy the time-dependent
Schr\"{o}dinger equation,
\re{
    i\h d|\psi(t)\ra/dt = (|c\qvp|^2 + m^2c^4)^\hf|\psi(t)\ra.
}{1}
In momentum representation the relativistic free-particle energy oper%
ator $(|c\qvp|^2 + m^2c^4)^\hf$ is transparently diagonal, but in con%
figuration representation it is a nonlocal integral operator.  This
was regarded as concerning by early quantum mechanics pioneers,
\e{not} for any \e{physical} reason, but because they feared it would
pose unpalatable \e{calculational} issues.  Perhaps they were not
mindful, in this regard, that relativistic corrections to the hydrogen
atom are expected to be of order one percent, and that Schr\"{o}dinger
had invented the bound state perturbation theory.  That notwith%
standing, Klein, Gordon and Schr\"{o}dinger decided to \e{iterate}
Eq.~(1) in order to rid it of its ``vexing'' square root, which maneu%
ver produces the Klein-Gordon equation~[1],
\re{
    (c^{-2}d^2/dt^2 + |\qvp/\h|^2 + \mu^2)|\psi(t)\ra = 0,
}{2}
where $\mu\eqdf((mc)/\h)$.  In configuration representation, where
$\qvp = -i\h\del$, it follows that $|\qvp/\h|^2 = -\del^2$, so
that aside from its mass term, the Klein-Gordon equation matches the
classical wave equation.  For every solution of Eq.~(1) that has a
definite momentum and positive energy, Eq.~(2) has another, \e{com%
pletely extraneous}, partner solution of the same momentum, but ener%
gy of the \e{opposite sign}.  Each such pair of energy-partner solu%
tions \e{fail} to be mutually orthogonal because they have the same
momentum, but this nonorthogonality of solutions with two different
energies contradicts a fundamental characteristic of quantum theory
that is necessary for its probability interpretation.  It is thus not
surprising that the Klein-Gordon theory gives rise to unacceptable
negative probabilities~[1].  We have seen that this is related to its
negative-energy solutions, which are not a feature of Eq.~(1).  The
associated negative energies as well have the physically problematic
trait of being \e{unbounded below}.  Since the Klein-Gordon equation
does not directly involve a Hamiltonian operator, but only the
\e{square} of such an operator, it cannot be related to the Heisen%
berg picture, the Heisenberg equations of motion or the Ehrenfest
theorem.  Thus it does not properly correspond to quantum mechanics
at all.  The straightforward and certainly physically most sensible
response to its defective nature is to unhesitatingly discard the
Klein-Gordon equation in favor of Eq.~(1), which is both \e{mandated}
by the correspondence principle and has \e{none} of the Klein-Gordon
equation's deficiencies.

Some physicists have been unaccountably reluctant to simply \e{heed}
these imperatives, and have cast about for a loophole which enables
the Klein-Gordon equation to survive.  Since the Klein-Gordon equation
abjectly fails the tests of quantum mechanics, the fact that it resem%
bles the classical wave equation has led to the notion that it too re%
presents a classical field theory.  To obtain any manner of quantum
physics from the Klein-Gordon equation, then, would presumably involve
quantizing the classical-field physics it represents.  The idea of
the Klein-Gordon equation as a classical field equation seems physi%
cally incongruous at first glance because that equation is associ%
ated with a \e{particle} which can perfectly well exist in a \e{va%
cuum}, whereas the only \e{familiar} classical field theory which does
\e{not} describe collective motions of \e{an underlying medium} is the
electromagnetic field.  In fact, by way of shedding light on this is%
sue of dynamical classical field theories which do \e{not} describe
motions of some medium, it has recently been shown that the classical
electric and magnetic fields which satisfy the source-free Maxwell
equations can be linearly mapped into the real and imaginary parts of
a transverse-vector wave function that, as a consequence, satisfies
the time-dependent Schr\"{o}dinger equation whose Hamiltonian operator
is $|c\qvp|$, which is, of course, physically appropriate to a mass%
less free particle, i.e., the free photon~[2].  Thus we have this el%
ectromagnetic example of \e{classical} field equations which in fact
are \e{equivalent} to a physically appropriately related \e{first-%
quantized Schr\"{o}dinger equation}---this equivalence has until
recently unfortunately been effectively hidden by the unfamiliar and
rather unusual linear mapping between the two equation systems.  If
\e{classical} Klein-Gordon field theory should \e{likewise} turn out
to be \e{physically appropriate quantum mechanics} that is \e{merely
disguised} by an unfamiliar linear mapping, then it would in fact be
entirely acceptable on that basis.  A caveat regarding this specula%
tion is of course that the correspondence principle \e{pinpoints} the
time-dependent Schr\"{o}dinger equation of Eq.~(1) as the \e{physical%
ly correct description of the quantum mechanics for this case of a
spinless relativistic free particle}.

An extremely important consideration regarding \e{classical} Klein-%
Gordon field theory is that the \e{quantum-mechanical} Eq.~(2) does
\e{not} define a \e{classical} Klein-Gordon field.  On physical mea%
surement grounds, a spinless \e{classical} field must be \e{strictly
real-valued in configuration representation}.  Therefore a \e{classi%
cal} Klein-Gordon field is a \e{real-valued function} $\phi(\vr, t)$
which satisfies the equation,
\re{
    (c^{-2}\pa^2/\pa t^2 - \del^2 + \mu^2)\phi(\vr,t) = 0.
}{3}
The fact that $\phi(\vr, t)$ is \e{real} will enable us to define a
conserved energy for this field that is \e{nonnegative}, which ef%
fectively \e{eliminates} the unphysical properties that flow from the
quantum-mechanical Klein-Gordon equation's \e{extraneous negative
energies}.

Since our \e{goal} for this classical Klein-Gordon field $\phi$ is its
\e{quantization}, we \e{must} cast it into \e{canonical Hamiltonian
form}, which involves a field $\pi$ that is canonically conjugate to
$\phi$, and also a corresponding conserved Hamiltonian from which
Hamilton's equations yield Eq.~(3).  The first step in this direction
is to find an action functional $S[\phi]$ that is stationary for those
$\phi$ which satisfy Eq.~(3).  This can in fact be obtained in terms
of a local \e{action density}, usually termed a Lagrangian density,
${\cal L}_\phi$ such that~[3],
\re{
    S[\phi] = \int{\cal L}_\phi\,d^3\vr\,dt.
}{4a}
For the classical Klein-Gordon field, ${\cal L}_\phi$ is conventional%
ly taken to be~[3],
\re{
{\cal L}_\phi = \hf(\dot\phi^2/c^2 - |\del\phi|^2 - \mu^2\phi^2),
}{4b}
which yields for the functional derivative of $S[\phi]$ with respect
to $\phi$,
\re{
\dl S[\phi]/\dl\phi = -\ddot\phi/c^2 + \del^2\phi - \mu^2\phi.
}{4c}
Setting this first-order variation of $S[\phi]$ with respect to $\phi$
to zero indeed produces the classical Klein-Gordon equation of
Eq.~(3).

With this Lagrangian density ${\cal L}_\phi$ in hand, we readily ob%
tain a field $\pi$ that is canonically conjugate to $\phi$~[3],
\re{
    \pi = \pa{\cal L}_\phi/\pa\dot\phi = \dot\phi/c^2,
}{5a}
which implies that,
\re{
    \dot\phi = c^2\pi.
}{5b}
The Hamiltonian density which corresponds to $\phi$ and $\pi$ is~[3],
\re{
{\cal H}_{\phi,\pi} = \lf.\dot\phi\pi - {\cal L}_\phi\rt|_{\dot\phi = c^2\pi}.
}{5c}
After combining this with Eq.~(4b), the result in terms of $\phi$ and
$\pi$ is,
\re{
{\cal H}_{\phi,\pi} = \hf(|\del\phi|^2 + \mu^2\phi^2 + c^2\pi^2),
}{5d}
which we note is indeed \e{nonnegative} for our \e{real-valued} clas%
sical $\phi$ and $\pi$ fields.  This \e{nonnegative} energy density is
a \e{crucial} feature of the \e{classical} Klein-Gordon field theory
with its \e{strictly real-valued fields}, as we have emphasized above.
From the Hamiltonian density of Eq.~(5d) and integration by parts we
obtain the Hamiltonian functional,
\re{
H[\phi, \pi] = \int{\cal H}_{\phi,\pi}\,d^3\vr =
            \hf\int\lf[\phi(-\del^2 + \mu^2)\phi + c^2\pi^2\rt]\,d^3\vr.
}{6a}
We now apply the standard prescriptions for Hamilton's equations of
motion to this Hamiltonian functional to obtain,
\re{
    \dot\phi = \dl H[\phi, \pi]/\dl\pi = c^2\pi,
}{6b}
and,
\re{
    \dot\pi = -\dl H[\phi, \pi]/\dl\phi = (\del^2 - \mu^2)\phi,
}{6c}
which together clearly imply the classical Klein-Gordon equation of
Eq.~(3).  Note that the canonically conjugate field $\pi$ satisfies
the Klein-Gordon equation as well.

Before going further with this now canonically Hamiltonized classical
Klein-Gordon field theory, we wish to digress in order to demonstrate
that, within a constant factor, \e{the real and imaginary parts of the
configuration representation of the state vector} $|\psi(t)\ra$ which
satisfies the Schr\"{o}dinger equation of Eq.~(1) are \e{also canoni%
cally conjugate classical fields} whose equations of motion follow
from the classical Hamiltonian functional which is given by the con%
served mean free-particle energy $\la\psi(t)|(|c\qvp|^2 + m^2c^4)^\hf|
\psi(t)\ra$.

\sct{Schr\"{o}dinger's equation from a classical Hamiltonian functional}
\noindent
It is apparent that the nonnegative conserved mean free-particle ener%
gy,
$$ \la\psi(t)|(|c\qvp|^2 + m^2c^4)^\hf|\psi(t)\ra, $$
is a nonnegative linear functional of \e{both} the wave function
$\psi(\vr, t)$ \e{and} its complex conjugate $\psi^\ast(\vr, t)$ when
it is expressed in the form,
\re{
    H[\psi, \psi^\ast] = \int\psi^\ast(\vr, t)
     \lf[(|c\qvp|^2 + m^2c^4)^\hf\,\psi\rt]\!\!(\vr,t)\; d^3\vr.
}{7a}
In configuration representation, the Hermitian operator $(|c\qvp|^2 +
m^2c^4)^\hf$ is a real, symmetric nonlocal \e{integral} operator whose
kernel is, of course, given by,
\re{
    \la\vr|(|c\qvp|^2 + m^2c^4)^\hf|\vr'\ra = (2\pi)^{-3}\int
    e^{i\vk\dt(\vr - \vr')}(|c\h\vk|^2 + m^2c^4)^\hf\,d^3\vk,
}{7b}
an integral whose result is distribution-valued, i.e., it is singular
as $|\vr - \vr'|\rta 0$, a feature which requires careful treatment
akin to that required by the delta function.   From Eq.~(7a) and
Eq.~(1) it is clear that time-dependent Schr\"{o}dinger equation for
the configuration-space wave function $\psi$ follows from the simple
functional differential equation,
\re{
    i\h\dot\psi = \dl H[\psi, \psi^\ast]/\dl\psi^\ast.
}{7c}

The complex-valued fields $\psi$ and $\psi^\ast$ have the dimensions
of probability density amplitude.  From these we readily define two
\e{strictly real-valued fields} $\phs$ and $\pis$ which each have the
dimensions of action density amplitude that is appropriate to their
being canonically conjugate,
\re{
      \phs\eqdf   (\h/2)^\hf(\psi + \psi^\ast),\qquad
      \pis\eqdf -i(\h/2)^\hf(\psi - \psi^\ast).
}{8a}
In terms of $\phs$ and $\pis$ we have that,
\re{
      \psi = (\phs + i\pis)/(2\h)^\hf,\qquad
 \psi^\ast = (\phs - i\pis)/(2\h)^\hf,
}{8b}
which when substituted into Eq.~(7a), bearing in mind that $\mu = ((m
c)/\h)$ and that in configuration representation $\qvp = -i\h\del$,
yields,
\re{
    H[\phs, \pis] = (c/2)\int\lf[\phs(-\del^2 + \mu^2)^\hf\,\phs +
                              \pis(-\del^2 + \mu^2)^\hf\,\pis\rt]\,d^3\vr.
}{9a}
Since,
\de{
    \dl H[\psi, \psi^\ast]/\dl\psi^\ast =
    (\dl H[\phs, \pis]/\dl\phs)(\pa\phs/\pa\psi^\ast) +
    (\dl H[\phs, \pis]/\dl\pis)(\pa\pis/\pa\psi^\ast),
}
we can readily obtain the real and imaginary parts of both the left
and right hand sides of Eq.~(7c) by application of Eqs.~(8a) and (8b).
The results of carrying out this decomposition are the two prescrip%
tions for equations of motion,
\re{
    \dot\phs =  \dl H[\phs, \pis]/\dl\pis,\qquad
    \dot\pis = -\dl H[\phs, \pis]/\dl\phs,
}{9b}
which are identical to the standard prescriptions for Hamilton's equa%
tions of motion.  This demonstrates that $\phs$ and $\pis$ are \e{in%
deed} canonically conjugate fields which pertain to the Hamiltonian
functional $H[\phs, \pis]$.  It is further readily verified by apply%
ing Eqs.~(8a) and (8b) that the prescriptions for Hamilton's equations
of motion of Eq.~(9b) for $\phs$ and $\pis$ are indeed \e{equivalent}
to Eq.~(7c), and therefore \e{are equivalent as well} to the time-de%
pendent Schr\"{o}dinger equation of Eq.~(1).  Actual application of
the prescriptions of Eq.~(9b) for Hamilton's equations of motion to
the Hamiltonian functional $H[\phs, \pis]$ of Eq.~(9a) yields,
\re{
    \dot\phs =  c(-\del^2 + \mu^2)^\hf\,\pis,\qquad
    \dot\pis = -c(-\del^2 + \mu^2)^\hf\,\phs.
}{9c}
From Eq.~(9c) and the first equation of Eq.~(8b), we readily show
that,
\re{
     i\h\dot\psi =  \h c(-\del^2 + \mu^2)^\hf\,\psi,
}{9d}
which, since $\mu = ((mc)/\h)$ and $\qvp = -i\h\del$ in configuration
representation, is, of course, equivalent to Eq.~(1).

It is therefore very clear indeed that the time-dependent Schr\"{o}%
dinger equation of Eq.~(1), which describes the solitary spinless rel%
ativistic free particle, is \e{equivalent} to the \e{classical Hamil%
tonian field system} that is described by the classical Hamiltonian
functional $H[\phs, \pis]$ of Eq.~(9a).  A very notable feature of
$H[\phs, \pis]$ is that it exhibits complete symmetry under the \e{in%
terchange} of its canonically conjugate fields $\phs$ and $\pis$; in%
deed these two fields as well have exactly the \e{same dimensions},
namely that of action density amplitude.  We now \e{return} to the
conventional Hamiltonian functional $H[\phi, \pi]$ of Eq.~(6a) for the
\e{classical} Klein-Gordon field $\phi$ and its canonical conjugate
$\pi$.

\sct{Canonical transformation of the classical Klein-Gordon field}
\noindent
We note that the \e{form} of the Hamiltonian functional of Eq.~(6a) is
\e{nonsymmetrical} under the \e{interchange} of $\phi$ and $\pi$; in%
deed $\phi$ and $\pi$ \e{themselves} have \e{different dimensions}.
It is straightforward to utilize fractional powers of the real, sym%
metric nonnegative operator $(-\del^2 + \mu^2)$ to devise a \e{canoni%
cal transformation} to \e{new} canonical fields that occur \e{symme%
trically} in the Hamiltonian functional, and which \e{both} have the
dimensions of action density amplitude.
This \e{canonical transformation} is given by,
\re{
    \phs = c^{-\hf}(-\del^2 + \mu^2)^\qr\,\phi,\qquad
    \pis = c^\hf(-\del^2 + \mu^2)^{-\qr}\,\pi,
}{10a}
which, of course, implies that,
\re{
    \phi = c^\hf(-\del^2 + \mu^2)^{-\qr}\,\phs,\qquad
    \pi = c^{-\hf}(-\del^2 + \mu^2)^\qr\,\pis.
}{10b}
Upon substituting these expressions for $\phi$ and $\pi$
in terms of their \e{canonical transforms} $\phs$ and $\pis$ into
the Hamiltonian of Eq.~(6a), and also taking note of the symmetric-op%
erator nature of powers of the operator $(-\del^2 + \mu^2)$, the new
Hamiltonian functional $H[\phs, \pis]$ is seen to exhibit complete
symmetry under the \e{interchange} of $\phs$ and $\pis$,
\re{
    H[\phs, \pis] = (c/2)\int\lf[\phs(-\del^2 + \mu^2)^\hf\,\phs +
                              \pis(-\del^2 + \mu^2)^\hf\,\pis\rt]\,d^3\vr.
}{10c}
Indeed this \e{canonically transformed} Hamiltonian functional of the
\e{classical} Klein-Gordon field theory is \e{identical} to that of
Eq.~(9a), whose equations of motion are, as we have noted above,
\e{entirely equivalent} to the time-dependent Schr\"{o}dinger equation
of Eq.~(1) for the solitary spinless relativistic free particle.  Thus
we have demonstrated the canonical equivalence of classical Klein-Gor%
don field theory to correspondence-principle first quantization of the
spinless relativistic free particle.

We can, in particular, provide the precise linear mapping of the real%
-valued classical Klein-Gordon field $\phi$ and its canonical conju%
gate $\pi$ into the complex-valued Schr\"{o}dinger wave function
$\psi$,
\re{
    \psi = (2\h c)^{-\hf}(-\del^2 + \mu^2)^\qr\,\phi +
           i(c/(2\h))^\hf(-\del^2 + \mu^2)^{-\qr}\,\pi,
}{11a}
which mapping can, of course, be inverted,
\re{
    \phi = ((\h c)/2)^\hf(-\del^2 + \mu^2)^{-\qr}(\psi + \psi^\ast),\qquad
    \pi = -i(\h/(2c))^\hf(-\del^2 + \mu^2)^\qr(\psi - \psi^\ast).
}{11b}
One can express the linear mapping of Eqs.~(11) \e{entirely} in terms
of the classical Klein-Gordon field $\phi$ \e{alone} by recalling from
Eq.~(5a) that $\pi = \dot\phi/c^2$.  It then becomes an interesting
exercise, which we leave to the reader, to verify that merely because
$\phi$ satisfies the Klein-Gordon equation, which is second-order in
time, the complex-valued wave-function construct $\psi$ of Eq.~(11a)
satisfies the Schr\"{o}dinger equation of Eq.~(1), which is first-%
order in time!  Nor is that all: the wave function construct $\psi$ of
Eq.~(11a) has also been painstakingly composed to have the dimensions
of probability density amplitude, which is appropriate to a wave
function; in fact its detailed construction is such that when it is
inserted into the Hamiltonian functional $H[\psi, \psi^\ast]$ of Eq.~%
(7a), which is the first-quantized free particle's conserved mean en%
ergy, the result is the Hamiltonian functional $H[\phi, \pi]$ of Eq.~%
(6a), which is, naturally enough, the energy of the corresponding
\e{classical} Klein-Gordon field.  The fact that a first-quantized
\e{free} particle's mean energy is \e{nonnegative}, along with the
\e{form} of the Klein-Gordon classical field's energy density that is
given by Eq.~(5d), together make it abundantly clear that the Klein-%
Gordon classical field is \e{restricted to being real-valued}, which
gives us additional insight into the \e{completely unphysical nature}
of the complex-valued quantum-mechanical Klein-Gordon wave function of
Eq.~(2). An alternative approach to verification of the Schr\"{o}%
dinger equation of Eq.~(1) from Eq.~(11a) is, of course, to apply the
Hamilton's equations of motion for $\phi$ and $\pi$ that are given by
Eqs.~(6b) and (6c).

\sct{Classical field quantization via particle second quantization}
\noindent
The quantization of the classical Klein-Gordon field is by far best
done in the Schr\"{o}dinger wave-function picture, where it is merely
\e{second quantization} of the spinless relativistic free particle,
which makes both the physics and the mathematics completely transpar%
ent.  This second quantization is achieved in the standard canonical
fashion by promoting the wave function $\psi$ and its complex conju%
gate $\psi^\ast$ to become the \e{Hermitian conjugate operators}
$\q\psi$ and $\q\psi^\dg$ which have the commutation relation~[3],
\re{
    [\q\psi(\vr), \q\psi^\dg(\vr')] = \dl^{(3)}\!(\vr - \vr').
}{12a}
With that, $\q\psi^\dg(\vr)$ is interpreted as the operator which
creates a particular type of spinless relativistic free particle of
mass $m$ at the position $\vr$, while $\q\psi(\vr)$ is interpreted
as the operator which annihilates such a particle at the position
$\vr$.  Fourier transforms of these operators perform these same
creation/annihilation functions in wave-vector (i.e., momentum)
space.

The real Hamiltonian functional $H[\psi, \psi^\ast]$ of Eq.~(7a) is
thereby quantized as the \e{Hermitian} operator $\qH[\q\psi, \q\psi%
^\dg]$, which is then taken to be the Hamiltonian operator of the sec%
ond-quantized spinless relativistic free particle system (or quantized
Klein-Gordon field system).  In the Heisenberg picture which is de%
fined by this Hamiltonian operator, the relativistic free-particle
time-dependent Schr\"{o}dinger equation of Eq.~(1) continues to hold
as a field-operator relation, i.e.,
\re{
    i\h\pa\q\psi(\vr, t)/\pa t =
                \h c\lf[(-\del^2 + \mu^2)^\hf\,\q\psi\rt]\!\!(\vr, t).
}{12b}

By multiple applications of the creation operator, arbitrarily large
numbers of this particular type of spinless relativistic free parti%
cle can be produced.  Indeed the Hermitian operator $\int\q\psi^\dg
\q\psi\,d^3\vr$ has the interpretation of particle number opera%
tor~[3].  The vast underlying Hilbert space is called Fock space~[3].

The relations of Eqs.~(11) may be simply transcribed into this se%
cond-quantized regime by noting the $\phi$ and $\pi$ become the
\e{Hermitian operators} $\q\phi$ and $\q\pi$, and of course $\psi$ and
$\psi^\ast$ become the operators $\q\psi$ and $\q\psi^\dg$ with their
fundamental canonical commutation relation and interpretation des%
cribed above.  These transcribed relations of Eqs.~(11) capture the
technical essence of the relationship between the quantized Klein-Gor%
don field theory and the second-quantized spinless relativistic free
particle picture that, because of the just-mentioned interpretation of
the precise role of the operators $\q\psi$ and $\q\psi^\dg$, is obvi%
ously vastly more physically and calculationally transparent.

\sct{Radiation-gauge vector potential equivalence to the first-quantized
     free photon}
\noindent
Maxwell's source-free equations for the transverse electric and
and magnetic fields have previously been canonically Hamiltonized by
a linear mapping that did not mix these fields, did but treat them in
a highly symmetric manner~[2].  This symmetric mapping of the source-%
free electric and magnetic fields into canonically conjugate counter%
parts turns out to be directly related to the real and imaginary parts
of the free photon's first-quantized transverse-vector wave function.
The traditional prescription for canonically Hamiltonizing source-free
electromagnetism in order to quantize it, on the other hand, invar%
iably eschews the electric and magnetic fields in favor of the vector
potential~[4], and therefore strongly parallels the treatment we have
just presented of the classical Klein-Gordon field.  The transverse
part of the electromagnetic vector potential $\vA_T$ satisfies the two
equations,
\re{
    \del\dt\vA_T = 0, \qquad\quad \ddot\vA_T/c^2 - \del^2\vA_T = \vj_T/c,
}{13a}
the second of which becomes simply the classical wave equation in the
source-free case.  In that case one may \e{also} use the \e{radiation
gauge}, for which $A^0 = 0$ and $\del\dt\vA = 0$~[4], conditions which
imply that the \e{only} nonvanishing part of the electromagnetic
four-potential $A^\mu$ is precisely $\vA_T$, and it, of course, satis%
fies the classical wave equation, which is \e{merely a special case of
the Klein-Gordon equation}.  For that reason one arrives at a Lagran%
gian density which strongly resembles the Lagrangian density of
Eq.~(4b),
\re{
    {\cal L}_{\vA} = \hf(|\dot\vA|^2/c^2 - |\del\x\vA|^2),
}{13b}
where $\vA$ is, of course, constrained to be \e{transverse}, i.e.,
$\del\dt\vA = 0$ and $\del\dt\dot\vA = 0$.  This Lagranian density
yields the canonical momentum $\vPi_A = \dot\vA/c^2$, and therefore
we have that $\dot\vA = c^2\vPi_A$.  Clearly $\vPi_A$ is \e{also}
transverse, i.e., $\del\dt\vPi_A = 0$.  With these results for the
canonical momentum, the corresponding Hamiltonian density comes out
to be,
\re{
    {\cal H}_{\vA, \vPi_A} = \hf(|\del\x\vA|^2 + c^2|\vPi_A|^2),
}{13c}
which is nonnegative and strongly resembles the Hamiltonian density
of Eq.~(5d).  Of course the Hamiltonian functional is the integral
of the Hamiltonian density over three-dimensional space.  Noting that
$\del\dt\vA = 0$ and integrating by parts, one readily puts the Hamil%
tonian functional into a form which is in essence the same as that of
Eq.~(6a) for the classical Klein-Gordon field,
\re{
 H[\vA, \vPi_A] = \hf\int\lf[\vA\dt(-\del^2\vA) + c^2|\vPi_A|^2\rt]\,d^3\vr.
}{14a}
This Hamiltonian functional has corresponding Hamilton's equations of
motion that are very similar to those of Eqs.~(6b) and (6c),
\re{
    \dot\vA = c^2\vPi_A,\qquad \dot\vPi_A = \del^2\vA.
}{14b}
Eq.~(14b) implies that both $\vA$ and $\vPi_A$ \e{obey the classical
wave equation}.

The nonsymmetry of $H[\vA, \vPi_A]$ under interchange of $\vA$ and
$\vPi_A$ now motivates a \e{canonical transformation} which is com%
pletely analogous to the one made in Eq.~(10a) for the classical
Klein-Gordon field,
\re{
    \vPhi = c^{-\hf}(-\del^2)^\qr\,\vA,\qquad
    \vPi = c^\hf(-\del^2 )^{-\qr}\,\vPi_A.
}{15a}
We note that both $\vPhi$ and $\vPi$ are transverse vector fields,
i.e., $\del\dt\vPhi = \del\dt\vPi = 0$.  This canonical transformation
results in a change to the form of the Hamiltonian functional, with a
result that is completely analogous to the Hamiltonian functionals of
Eqs.~(10c) and (9a),
\re{
   H[\vPhi, \vPi] = \hf\int\lf[\vPhi\dt\lf(c(-\del^2)^\hf\,\vPhi\rt) +
                            \vPi\dt\lf(c(-\del^2)^\hf\,\vPi\rt)\rt]\,d^3\vr.
}{15b}
The Hamiltonian functional $H[\vPhi, \vPi]$ has corresponding Hamil%
ton's equations of motion that are very similar to those of Eq.~(9c),
\re{
 \dot\vPhi = c(-\del^2)^\hf\,\vPi,\qquad \dot\vPi = -c(-\del^2)^\hf\,\vPhi.
}{15c}

If one now, in analogy with Eq.~(8b), defines,
\re{
    \vPsi\eqdf(\vPhi + i\vPi)/(2\h)^\hf,
}{16a}
which has the dimensions of probability density amplitude appropriate
to a wave function, one deduces from Eq.~(15c) that its equation of
motion is,
\re{
    i\h\dot\vPsi = \h c(-\del^2)^\hf\,\vPsi,
}{16b}
which is precisely that of the time-dependent Schr\"{o}dinger equation
for the free photon because, since $\qvp = -i\h\del$ in configuration
representation, $\h c(-\del^2)^\hf = |c\qvp|$ in that representation.
We further note that this complex-valued free-photon wave function
$\vPsi$ of Eq.~(16a) is of course a transverse vector field, i.e.,
$\del\dt\vPsi = 0$, and that its dimensions are indeed appropriate to
a wave function.

We have thus demonstrated that in the case of source-free electromag%
netism there exists a linear mapping of the electromagnetic trans%
verse-vector canonically conjugate fields $\vA$ and $\vPi_A$ into the
real and imaginary parts of the free photon wave function $\vPsi$.
This linear mapping is, of course, highly analogous to that given in
Eq.~(11a) for the classical Klein-Gordon field,
\re{
    \vPsi = (2\h c)^{-\hf}(-\del^2)^\qr\,\vA +
           i(c/(2\h))^\hf(-\del^2)^{-\qr}\,\vPi_A,
}{17a}
and its inverse is,
\re{
    \vA = ((\h c)/2)^\hf(-\del^2)^{-\qr}(\vPsi + \vPsi^\ast),\qquad
    \vPi_A = -i(\h/(2c))^\hf(-\del^2)^\qr(\vPsi - \vPsi^\ast).
}{17b}
The Schr\"{o}dinger equation of Eq.~(16b) and the result for $\vA$
given by Eq.~(17b) are exactly the same as those which were previously
obtained via the symmetric canonical Hamiltonization of the electric
and magnetic fields which satisfy Maxwell's source-free equations~[2].

One can express the linear mapping of Eqs.~(17) \e{entirely} in terms
of the transverse vector potential $\vA$ \e{alone} by recalling that
$\vPi_A = \dot\vA/c^2$ (e.g., see Eq.~(14b)).  It then becomes an in%
teresting exercise to verify that the time-dependent Schr\"{o}dinger
equation of Eq.~(16b) for the free photon's wave function $\vPsi$, as
given by Eq.~(17a) and supplemented by the formula $\vPi_A = \dot\vA/
c^2$, follows \e{merely} from the fact that the transverse vector
potential $\vA$ satisfies the \e{classical wave equation}, notwith%
standing that the classical wave equation is second-order in time,
whereas the Schr\"{o}dinger equation of Eq.~(16b) is first-order in
time!  Of course an alternative approach to verification of the
Schr\"{o}dinger equation of Eq.~(16b) from Eq.~(17a) is to apply the
Hamilton's equations of motion for $\vA$ and $\vPi_A$ that are given
by Eq.~(14b).

\sct{Conclusion}
\noindent
The most striking results of this paper are the one-to-one linear map%
pings, given by Eqs.~(11) and (17), of real-valued dynamical classical
fields that are described by simple, second-order in time, wave equa%
tions, i.e., the classical Klein-Gordon equation and the classical
wave equation, onto correspondence-principle first-quantized relati%
vistic Schr\"{o}dinger-equation wave functions for free particles,
i.e., spinless relativistic free particles and free photons---particu%
larly in light of the fact that such correspondence-principle first-%
quantized free-particle wave functions can be transparently converted
in a flash into the free-particle annihilation and creation operators
that are the very heart of the quantum many-free-particle description.
These one-to-one linear mappings lend a truly gratifying dollop of
theoretical concreteness to the heretofore vague notion of \e{comple%
mentarity} between dynamical classical wave fields and the correspon%
ding particle quanta.

Fascinating though it is that classical Klein-Gordon field theory is
equivalent to the elementary correspon\-dence-principle first quanti%
zation of the spinless relativistic free particle, there is in fact no
practical point to the exercise.  Elementary correspondence-principle
first quantization is clearly enormously simpler and more physically
transparent than is long-winded canonical Hamiltonization and canoni%
cal transformation of classical field equations that are second-order
in time.  Furthermore, elementary correspondence-principle \e{first}
quantization is by far the most physically and mathematically trans%
parent \e{gateway} to field quantization because its \e{second} quan%
tization is so extraordinarily straightforward and simple.  It is only
for electromagnetism, where classical field theoretic methods have
been completely entrenched for a century and a half, that discussion
of the relationship of classical field theory to elementary correspon%
dence-principle first quantization is at all worthwhile.  Even there,
field quantization definitely ought to be handled by proceeding di%
rectly from the photon's elementary correspondence-principle first
quantization to its second quantization---there is clearly no alterna%
tive field-theory approach that is as physically transparent or calcu%
lationally simple.

There can by now be no doubt at all that elementary correspondence-%
principle first- and second-quan\-tization is physically correct in
all cases~[2, 5].  It has clearly proved its mettle vis-\`a-vis both
classical electromagnetism and classical Klein-Gordon field theory.
As for Dirac theory, it is even \e{more} unphysical (if such a thing
is possible!) than is treating the classical Klein-Gordon field as a
complex-valued wave function.  Not merely does Dirac theory manifest
the signature unbounded-below energies, its artificially introduced
anticommuting matrices cause the commutators of some \e{very basic ob%
servables} to behave insanely.  The commutator of any two \e{orthogon%
al} components of the free-particle Dirac \e{velocity operator} not
only \e{fails} to vanish, its value \e{is not affected in the slight%
est} by taking the \e{classical limit} $\h\rta 0$.  And when the
\e{nonrelativistic limit} $c\rta\infty$ is taken, these orthogonal
velocity-component commutators \e{diverge}!  The \e{free-particle} Di%
rac theory as well violates Newton's first law of motion with mind-%
boggling spontaneous particle \e{acceleration} that is \e{inversely}
proportional to Planck's constant and directly proportional to parti%
cle mass and the cube of the speed of light; it also features a \e{un%
iversal fixed particle speed} which is 73\% \e{greater} than that of
light, and it unaccountably manifests strong spontaneous spin-orbit
coupling~[5].  In light of the fact that elementary correspondence-%
principle first- and second-quantization has \e{no discernible pathol%
ogies whatsoever}, it is patently obvious what the physics status of
Dirac theory ought to be.

\vskip 1.75\baselineskip\noindent{\frtbf References}
\vskip 0.25\baselineskip

{\parindent = 15pt
\sk\item{[1]}
J. D. Bjorken and S. D. Drell,
\e{Relativistic Quantum Mechanics}
(McGraw-Hill, New York, 1964).
\sk\item{[2]}
S. K. Kauffmann,
arXiv:1011.6578 [physics.gen-ph]
(2010).
\sk\item{[3]}
S. S. Schweber,
\e{An Introduction to Relativistic Quantum Field Theory}
(Harper \& Row, New York, 1961).
\sk\item{[4]}
J. D. Bjorken and S. D. Drell,
\e{Relativistic Quantum Fields}
(McGraw-Hill, New York, 1965).
\sk\item{[5]}
S. K. Kauffmann,
arXiv:1009.3584 [physics.gen-ph]
(2010).
}
\bye